%% file: 00_Main.tex
\documentclass[sigconf]{acmart}
\AtBeginDocument{%
  }

\copyrightyear{2025}
\acmYear{2025}
\setcopyright{cc}
\setcctype{by}
\acmConference[CIKM '25]{Proceedings of the 34th ACM International
Conference on Information and Knowledge Management}{November 10--14,
2025}{Seoul, Republic of Korea}
\acmBooktitle{Proceedings of the 34th ACM International Conference on
Information and Knowledge Management (CIKM '25), November 10--14, 2025,
Seoul, Republic of Korea}
\acmDOI{10.1145/3746252.3761471}
\acmISBN{979-8-4007-2040-6/2025/11}

\begin{document}

\title{SQuAI: Scientific Question-Answering with Multi-Agent Retrieval-Augmented Generation}

\author{Ines Besrour}
\authornote{Both authors contributed equally to this research.}
\email{ines.besrour@mailbox.tu-dresden.de}
\orcid{0009-0007-2736-8971}

\author{Jingbo He}
\authornotemark[1]
\email{jingbo.he@mailbox.tu-dresden.de}
\orcid{0009-0000-3653-2840}

\affiliation{%
  \institution{TU Dresden, ScaDS.AI}
  \city{Dresden}
  \country{Germany}
}

\author{Tobias Schreieder}
\email{tobias.schreieder@tu-dresden.de}
\affiliation{%
  \institution{TU Dresden, ScaDS.AI}
  \city{Dresden}
  \country{Germany}}
\orcid{0009-0000-8268-4204}

\author{Michael Färber}
\email{michael.faerber@tu-dresden.de}
\affiliation{%
  \institution{TU Dresden, ScaDS.AI}
  \city{Dresden}
  \country{Germany}}
\orcid{0000-0001-5458-8645}

\renewcommand{\shortauthors}{Ines Besrour, Jingbo He, Tobias Schreieder, and Michael Färber}

\begin{abstract}
We present SQuAI (\url{https://squai.scads.ai/}), a scalable and trustworthy multi-agent retrieval-augmented generation (RAG) framework for scientific question answering (QA) with large language models (LLMs). SQuAI addresses key limitations of existing RAG systems in the scholarly domain, where complex, open-domain questions demand accurate answers, explicit claims with citations, and retrieval across millions of scientific documents. Built on over 2.3 million full-text papers from arXiv.org, SQuAI employs four collaborative agents to decompose complex questions into sub-questions, retrieve targeted evidence via hybrid sparse-dense retrieval, and adaptively filter documents to improve contextual relevance. To ensure faithfulness and traceability, SQuAI integrates in-line citations for each generated claim and provides supporting sentences from the source documents. Our system improves faithfulness, answer relevance, and contextual relevance by up to +0.088 (12\%) over a strong RAG baseline. We further release a benchmark of 1,000 scientific question–answer–evidence triplets to support reproducibility. With transparent reasoning, verifiable citations, and domain-wide scalability, SQuAI demonstrates how multi-agent RAG enables more trustworthy scientific QA with LLMs.
\end{abstract}



\maketitle

\input{01_Introduction}
\input{02_Related_Work}
\input{03_Methodology}
\input{04_Evaluation}
\input{05_Conclusion}

\begin{acks}
We thank Norman Koch for his support in setting up the demo system on the compute resources provided by the high-performance computing center at the NHR Center of TU Dresden.

The authors acknowledge the financial support by the Federal Ministry of Research, Technology and Space of Germany and by Sächsische Staatsministerium für Wissenschaft, Kultur und Tourismus in the programme Center of Excellence for AI-research „Center for Scalable Data Analytics and Artificial Intelligence Dresden/Leipzig“, project identification number: ScaDS.AI.
\end{acks}

\section*{GenAI Usage Disclosure}
The authors utilized GPT-4o for grammar and spelling checks, minor rewording, and structural editing support. The AI tools did not contribute to the intellectual content or scientific conclusions. All content was subsequently reviewed and edited by the authors, who assume full responsibility for the publication.

\bibliographystyle{ACM-Reference-Format}
\bibliography{literature}

\end{document}

%% file: 01_Introduction.tex
\section{Introduction}
\label{introduction}

Large Language Models (LLMs) have shown strong performance on tasks such as multi-step reasoning~\cite{Wei2022CoT}, summarization~\cite{Zhang2024Summarization}, and open-domain question answering (QA)~\cite{Singhal2025QA}. Nonetheless, LLMs often generate hallucinations, i.e., factually incorrect or misleading information~\cite{Huang2025Hallucination}. One approach to mitigating hallucinations is retrieval-augmented generation (RAG), which combines traditional information retrieval with LLMs~\cite{Lewis2020RAG}. In RAG, relevant documents are retrieved from external sources and used to ground the model’s responses in curated, up-to-date information while reducing the likelihood of unsupported claims.

Scientific QA is a challenging setting, where hallucinations can have especially severe consequences. Unlike general-domain QA, it requires not only factually correct answers but also precise scientific terminology, long-form reasoning, and integration of evidence from multiple sources. Although RAG reduces hallucinations by grounding answers in retrieved documents, standard RAG methods treat questions monolithically and retrieve in a single step, often yielding incomplete or only marginally relevant evidence. Verifiability of LLM-generated answers is also crucial, and as proposed by \citet{Gao2023ALCE}, fine-grained in-line citations help users trace claims back to their sources and assess their reliability.

To overcome the limitations of standard RAG in scientific QA, we present \textit{SQuAI}, a collaborative multi-agent RAG system. Our system is built on the newly released \textit{unarXive 2024} dataset~\cite{Besrour2025UnarXive}, which updates previous versions~\cite{Saier2020unarXive, Saier2024unarxive} to provide large-scale, open-domain coverage of all publications from arXiv, enabling QA across a wide range of scientific topics. SQuAI provides an end-to-end QA user interface (UI) where users can select the LLM backend, configure retrieval systems such as sparse, dense, or hybrid, and examine intermediate reasoning steps during generation. 

In contrast to existing systems like BioRAGent~\cite{Ateia2025BioRAGent}, PaperQA~\cite{Lála2023PaperQA}, and ScienceQA~\cite{Pan2022ScienceQA}, SQuAI supports a wide range of scientific domains and offers adaptability to different use cases through configurable retrieval and generation settings. BioRAGent is restricted to biomedical QA, relies on static query expansion with snippet-level citation, and provides limited control over retrieval and answer generation. PaperQA retrieves information from full-text scientific papers but is designed primarily for offline use and lacks interactive, real-time capabilities. ScienceQA, while addressing scientific QA, does not generate in-line citations and is not based on a multi-agent architecture, limiting its traceability compared to SQuAI.

SQuAI addresses key limitations of existing scientific QA systems through four core innovations. First, it achieves \textit{high answer relevancy} via a multi-agent architecture that decomposes user questions into sub-questions to enable more accurate evidence aggregation than standard RAG approaches. Second, it ensures \textit{contextual relevancy} through adaptive document filtering, which prioritizes pertinent content for complex, multi-faceted queries, and through hybrid retrieval that combines sparse and dense approaches to improve literature coverage. Third, it improves \textit{faithfulness} through traceability by providing fine-grained in-line citations along with citation context, i.e., the exact sentences supporting each claim, thus enabling transparent verification. Finally, SQuAI is designed to be \textit{scalable} for real-world usage, operating on the full-text corpus of over 2.3 million scientific publications across multiple disciplines including computer science, mathematics, and physics. To our knowledge, SQuAI is the first publicly available system to integrate these capabilities in a unified platform for real-time, verifiable scientific QA at scale. SQuAI is accessible online,\footnote{\url{https://squai.scads.ai/}} and the accompanying code and datasets are available in our repository.\footnote{\url{https://github.com/faerber-lab/SQuAI}}

Our contributions can be summarized as follows:

\begin{itemize}
\item We present SQuAI, a multi-agent RAG system for scientific QA over all arXiv publications up to 2024.
\item We design a user-facing QA interface with modular configuration, transparent reasoning inspection, and fine-grained in-line citations with citation context for verification.
\item We release a novel synthetic dataset with 1,000 scientific question–answer–evidence triplets tailored to unarXive 2024, supporting evaluation of scientific QA systems.
\item We conduct an extensive evaluation, showing that SQuAI improves a combined score of Faithfulness and Answer / Contextual Relevancy by +0.088 (12\%) over a RAG baseline.
\end{itemize}

%% file: 02_Related_Work.tex
\section{Related Work}
\label{related_work}

\textbf{Multi-Agent Retrieval-Augmented Generation.} Recent RAG research increasingly adopts multi-agent systems to enhance QA tasks. \citet{Zhao2024Longagent} introduced \textit{LongAgent}, where sub-tasked agents handle document segments under a coordinating leader. \textit{IM-RAG} models an inner monologue between retriever, reasoner, and refiner agents to iteratively refine queries and results~\cite{Yang2024IM-RAG}, with related work exploring agent specialization for planning or verification~\cite{Jang2024AU-RAG, Saeid2024AgentFusion}. \citet{Chang2024MAIN-RAG} introduced \textit{MAIN-RAG}, an architecture that improves evidence selection under noisy retrieval. Our work builds upon MAIN-RAG, further improving retrieval and refining answer generation through a fourth agent for query decomposition and through fine-grained in-line citations.

\textbf{Attributed Text Generation.} Also known as citation generation, this task aims to produce text with explicit links to source documents used for text generation, enhancing trust and verifiability. As discussed by \citet{Huang2024Citation, Schreieder2025AttributionCitationQuotation}, attribution approaches include parametric methods, which rely solely on internal model knowledge like Galactica~\cite{Taylor2022Galactica}, and non-parametric methods that access external sources. Non-parametric approaches are further divided into \textit{post-generation}, where the model first generates an answer and then identifies supporting evidence~\cite{Gao2023RARR, Ramu2024PostHoc}, and \textit{post-retrieval}, which retrieve evidence before generation following the RAG framework~\cite{Gao2023ALCE, Fierro2024LearningToPlan, Berchansky2024CoTAR}. Our work advances post-retrieval approaches through a multi-agent architecture with query decomposition to retrieve more relevant evidence and generate a combined answer with fine-grained in-line citations and rich citation context.
 
\textbf{Scientific Question Answering.} Scientific QA tasks are typically categorized by input source and answer format. Input sources include \textit{open-domain} settings, such as SciQA~\cite{Auer2023SciQA} and LitSearch~\cite{Ajith2024LitSearch}, where answers are retrieved from corpora, and \textit{document-grounded} settings, such as Qasper~\cite{Dasigi2021Qasper}, where answers are derived from a provided document.
Answer formats vary widely. Some tasks use multiple-choice or yes/no questions, such as ScienceQA~\cite{Pan2022ScienceQA}, while others require short factoid or list-style responses, as in BioASQ-QA~\cite{Krithara2023BioASQ-QA} and PubMedQA~\cite{Jin2019PubMedQA}. Some questions require long-form, free-text explanatory answers, as seen in SciQA~\cite{Auer2023SciQA}.
While biomedical QA has advanced, \textit{open-domain} long-form scientific QA remains underexplored. To help address this gap, we leverage questions from LitSearch~\cite{Ajith2024LitSearch}, while using unarXive 2024~\cite{Besrour2025UnarXive} as retrieval corpus due to its broad and diverse coverage of scientific literature.

%% file: 03_Methodology.tex
\section{System Overview}
\label{methodology}

SQuAI is built on unarXive 2024~\cite{Besrour2025UnarXive}, a large-scale structured collection of all full-text arXiv papers from 1991 to 2024. Each paper includes rich metadata, annotated citations, section boundaries, LaTeX equations, and a detailed citation network. With substantial textual volume and dense technical content organized in deep section hierarchies, these papers provide a challenging benchmark for complex retrieval and citation tasks, enabling robust evaluation of RAG systems in realistic scientific QA settings.

\begin{figure*}[htbp]
  \centering
  \includegraphics[width=1\textwidth]{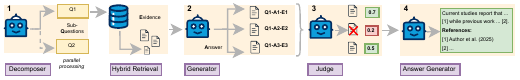}
  \Description{Visualization of the SQuAI framework for scientific QA.}
  \caption{An overview of the SQuAI framework including four agents and hybrid retrieval for scientific QA.}
  \label{fig:system_overview}
\end{figure*}

Adopting the multi-agent filtering approach from the MAIN-RAG framework~\cite{Chang2024MAIN-RAG}, we introduce a substantially enhanced multi-agent architecture for scientific QA. We introduce a novel query decomposition and generate precise in-line citations. In Figure~\ref{fig:system_overview}, we detail our agents’ roles: Agent-1 decomposes the input query into sub-questions. For each sub-question, a hybrid retriever selects top-k documents, Agent-2 generates initial answers, Agent-3 filters question-answer-evidence (Q-A-E) triplets. Finally, Agent-4 generates the final answer with in-line citations and citation context.

\subsection{Multi-Agent System Design}
The overall architecture comprises five main components:

\subsubsection{Agent-1: Decomposer}
Agent-1 serves as the initial stage of the QA pipeline, tasked with decomposing complex user queries into simpler, semantically distinct sub-questions. For example, the query \textit{“What is quantum computing and how is it used in cryptography?”} can be decomposed into two sub-questions:\\ \textit{“What is quantum computing?”} and \textit{“How is quantum computing used in cryptography?”}

This decomposition enables more precise retrieval for each aspect of the original query, enhancing SQuAI's evidence aggregation.
Complex or multi-faceted queries are common in real-world information-seeking tasks, particularly in scientific and technical domains~\cite{nogueira2020document, khattab2021relevance}. Prior work has emphasized the benefits of handling these cases via decomposition to improve both retrieval and downstream answer quality~\cite{wolfson2020break}.
Our system handles complex queries by decomposing them into sub-questions, enabling targeted retrieval and reducing ambiguity across large scientific corpora.
Subsequently, each sub-question is processed independently and in parallel by \textbf{Hybrid Retrieval, (2) Generator,} and \textbf{(3) Judge.}

\subsubsection{Hybrid Retrieval}
Our system combines sparse and dense retrieval models to maximize both lexical and semantic coverage. Motivated by prior work~\cite{rayo2025hybrid}, this hybrid approach aims to improve retrieval performance, particularly for complex scientific texts. For sparse retrieval, we use \textit{BM25}, a well-established exact term matching model~\cite{Robertson1994BM25, Robertson1994Okapi}, while dense retrieval relies on \textit{intfloat/e5-base-v2} embeddings~\cite{Wang2024E5}. We primarily use paper abstracts for indexing, supported by findings from the LitSearch paper~\cite{Ajith2024LitSearch} that full-text retrieval offers only marginal gains. Limiting retrieval scope is crucial for scalability across millions of long-scale documents without sacrificing performance. The final document scores are computed by interpolating both retrieval scores.
\begin{equation}
S_{\mathrm{hybrid}}(d) = \alpha \cdot S_{\mathrm{sparse}}(d) + (1 - \alpha) \cdot S_{\mathrm{dense}}(d)
\end{equation}
Following the findings of~\citet{rayo2025hybrid}, we set $\alpha = 0.35$ to slightly favor dense retrieval while preserving complementary signals from sparse retrieval.

\subsubsection{Agent-2: Generator} 
After document retrieval, we select up to $\texttt{top\_k}$ candidate papers per sub-query, where $\texttt{top\_k}$ is adjustable between 3 and 10 to balance the number of documents retrieved with processing efficiency. Agent-2 processes each document individually within its sub-query context to generate Q-A-E triplets, which serve as structured input for Agent 3. More than just creating Q-A pairs, this agent isolates potentially relevant content from each document, enabling a more fine-grained evaluation and selection of information in the next stage.

\subsubsection{Agent-3: Judge}
This agent evaluates each Q-A-E triplet to determine whether the document is relevant and supportive for answering the query. For each triplet, Agent-3 generates a binary assessment (“Yes” or “No”) in response to the question: “Is this document relevant for answering the query?”
To enable adaptive filtering beyond binary decisions, we compute a relevance confidence score (\textit{RelScore}) based on the model's output probabilities:
\[
\text{RelScore} = \log p(\text{``Yes''}) - \log p(\text{``No''})
\]
This scoring method allows us to distinguish strong vs. weak affirmatives and to dynamically adjust filtering sensitivity based on query difficulty.
Instead of a fixed threshold for filtering retrieved documents by relevance scores, we compute a query-specific threshold using the mean score ($\tau_q$) and standard deviation ($\sigma$) of all candidate documents. The system retains a document if:
\[
\text{RelScore} \geq \tau_q - n \cdot \sigma
\]
where $n$ is a hyperparameter that controls filtering stringency. Based on empirical results from the MAIN-RAG study, $n = 0.5$ yields the best trade-off between precision and recall~\cite{Chang2024MAIN-RAG}.

\subsubsection{Agent-4: Answer Generator}
Following document filtering, Agent-4 synthesizes a final, coherent answer for all sub-queries by combining the full-text from the relevant documents retained by Agent-3. Crucially, this step also requires precise citation placement: Each factual assertion must be followed by a citation in the standardized $[X]$ format, linking to the associated arXiv entry.

The model must identify discrete factual claims within the answer, determine which documents support each claim (which may vary in style or coverage), and insert the correct citation(s) immediately after. This includes handling multiple sources for a single claim (e.g., \texttt{[1][2]}) and integrating complementary or conflicting evidence across documents. To guide the model, we use a structured few-shot prompt with three carefully crafted exemplars illustrating correct and incorrect citation practices. This lightweight alignment encourages the model to generalize proper citation behavior without fine-tuning. The model is instructed that every factual sentence must include at least one citation and should reference diverse sources rather than relying heavily on a single document. The Answer Generator must balance informativeness with traceability by citing appropriate evidence or rephrasing/removing unsupported claims. This ensures the output is grounded in retrieved papers and meets academic standards of transparency and attribution. In addition Agent-4 is prompted to extract the citation context.

Below is an example of an appropriate answer generated in response to the previous question:
\textit{Quantum computing uses qubits to perform computations based on quantum mechanics [1]. It has potential applications in cryptography, particularly for breaking classical encryption schemes (...)[2].}

\begin{table*}[t!]
\centering
\caption{Evaluation of SQuAI on LitSearch, unarXive Simple, and Expert with metrics Answer Relevance (Ans.), Contextual Relevance (Con.), Faithfulness (Fai.), and their mean (Avg.). We compare standard RAG with SQuAI using abstracts and full-texts.}
\label{tab:evaluation_squai_hybrid}
\begin{tabular}{l|cccc|cccc|cccc}
\toprule
\multicolumn{1}{c|}{} & \multicolumn{4}{c|}{LitSearch} & \multicolumn{4}{c|}{unarXive Simple} & \multicolumn{4}{c}{unarXive Expert} \\
\cmidrule(lr){2-5} \cmidrule(lr){6-9} \cmidrule(lr){10-13}
Approach & Ans. & Con. & Fai. & Avg. & Ans. & Con. & Fai. & Avg. & Ans. & Con. & Fai. & Avg. \\
\midrule
Standard RAG               & 0.897 & 0.513 & 0.983 & 0.798 & 0.750 & 0.562 & 0.965 & 0.759 & 0.762 & 0.643 & 0.984 & 0.796 \\
SQuAI \textit{(Abstract)}  & 0.903 & \textbf{0.739} & 0.966 & \textbf{0.869} & 0.748 & \textbf{0.782} & 0.954 & 0.828 & 0.808 & \textbf{0.653} & 0.974 & 0.812 \\
SQuAI \textit{(Full Text)} & \textbf{0.937} & 0.677 & \textbf{0.984} & 0.866 & \textbf{0.883} & 0.670 & \textbf{0.988} & \textbf{0.847} & \textbf{0.948} & 0.649 & \textbf{0.995} & \textbf{0.864} \\
\bottomrule
\end{tabular}
\end{table*}

\subsection{User Interface}

SQuAI is implemented as a web application using the Streamlit framework for a clean and responsive UI, and FastAPI for the backend.
We offer several key features designed for transparency and flexibility. Users can enter a question via a dedicated input field and initiate retrieval with a search button. For added interpretability, the UI displays the decomposed sub-queries generated by SQuAI. Answers are shown in text boxes with in-line citations, accompanied by a list of reference snippets, each including direct, clickable links to the corresponding arXiv papers. Users can configure parameters that balance different retrieval methods and control strictness of document filtering. The UI also allows the selection of a retrieval model, such as hybrid, sparse, or dense, based on the task.

%% file: 04_Evaluation.tex
\section{Evaluation}
\label{evaluation}

To evaluate SQuAI’s performance, we employ \textit{DeepEval}~\cite{DeepEval2025}, an open-source LLM-as-a-judge framework that offers prompt-based metrics delivering continuous scores along with detailed supporting rationales. These evaluation metrics are implemented using chain-of-thought prompting inspired by G-Eval~\cite{Liu2023G-Eval}. We report results across three benchmarks, providing a thorough comparison of SQuAI’s performance against a state-of-the-art RAG baseline.

\subsection{Benchmarks \& Metrics}

We evaluate the system performance across three QA benchmarks focused on long-form scientific QA. All benchmarks are based on arXiv publications and designed to test different aspects of RAG across varying complexity levels.

\textbf{LitSearch.} This retrieval benchmark consists of literature-search questions derived from computer science papers~\cite{Ajith2024LitSearch}. The questions originate from both GPT-4-generated prompts and author-written queries. We use a subset of 478 questions that include at least one arXiv ID as ground truth, ensuring compatibility with unarXive.

\textbf{unarXive Simple.} This benchmark consists of 500 long-form, open-domain questions synthetically generated using DeepEval and LLaMa 3.3 70B Instruct. The questions are derived from individual papers and are designed to be less complex and more general, making them suitable for a broad, non-specialist audience.

\textbf{unarXive Expert.} Similar to unarXive Simple, this benchmark contains 500 questions generated using DeepEval and LLaMa 3.3 70B Instruct. The questions are more specific and technical, requiring detailed evidence from the source paper to answer.

We assess answer quality using three key DeepEval metrics. Each metric uses the LLaMA 3.3 70B Instruct model and structured reasoning steps modeled on G-Eval’s chain-of-thought methodology. For each Q-A-E triplet, the model outputs three scalar scores in \([0, 1]\), each with an intermediate rationale. We avoid evaluating against synthetic reference answers to prevent circularity, as both system and baseline outputs are LLM-generated. Instead, we focus on verifiability by judging answers directly against the evidence they cite, ensuring a traceable and source-grounded evaluation.

\textbf{Answer Relevancy.} This metric evaluates how well the answer addresses the given question, focusing on semantic alignment between the question and the generated response.

\textbf{Contextual Relevancy.} This metric measures how effectively the answer incorporates the provided evidence passages, assessing whether cited content is meaningfully used.

\textbf{Faithfulness.} This metric assesses whether the answer remains accurate with respect to evidence, penalizing unsupported claims.

\subsection{Results}
We compare SQuAI against standard RAG, which follows a conventional setup without multi-agent orchestration or query decomposition. The model is guided by a single prompt including task instructions and the retrieved papers to generate an answer.
Table \ref{tab:evaluation_squai_hybrid} shows that SQuAI, using hybrid retrieval, consistently improves upon the baseline across all benchmarks. While both settings yield gains, the abstract-only variant shows larger and more consistent improvements (+0.068 to +0.088) than the full-text setting (+0.016 to +0.071). Answer Relevance is higher for SQuAI (Full-Text), while Contextual Relevance is higher for SQuAI (Abstract), reflecting a trade-off where more context better addresses the question but can overwhelm the LLM with excessive information. These results highlight that abstract-based generation provides strong performance benefits with lower computational overhead, making it a more efficient choice for scientific QA in certain scenarios. Notably, SQuAI achieves large improvements over standard RAG in Answer Relevance and Contextual Relevance, while Faithfulness remains reliably high with scores exceeding 0.95 across all configurations.

%% file: 05_Conclusion.tex
\section{Conclusion}
\label{conclusion}

With SQuAI we introduced a multi-agent RAG framework to enhance trustworthiness in scientific QA. SQuAI outperforms a state-of-the-art RAG baseline by up to +0.088 (12\%) for \textit{faithfulness}, \textit{answer relevance}, and \textit{contextual relevance}. These gains arise from four key innovations. First, to improve \textit{answer relevance}, we decompose complex questions into sub-questions for more accurate evidence aggregation. Second, to enhance \textit{contextual relevance}, we apply adaptive document filtering and hybrid retrieval for broad and relevant literature coverage. Third, to strengthen \textit{faithfulness}, we provide fine-grained in-line citations and context for transparent verification. Finally, we ensure SQuAI scales to real-world use, operating over 2.3 million scientific papers across diverse domains.

%% file: 00_Main.bbl

\begin{thebibliography}{38}


\ifx \showCODEN    \undefined \def \showCODEN     #1{\unskip}     \fi
\ifx \showISBNx    \undefined \def \showISBNx     #1{\unskip}     \fi
\ifx \showISBNxiii \undefined \def \showISBNxiii  #1{\unskip}     \fi
\ifx \showISSN     \undefined \def \showISSN      #1{\unskip}     \fi
\ifx \showLCCN     \undefined \def \showLCCN      #1{\unskip}     \fi
\ifx \shownote     \undefined \def \shownote      #1{#1}          \fi
\ifx \showarticletitle \undefined \def \showarticletitle #1{#1}   \fi
\ifx \showURL      \undefined \def \showURL       {\relax}        \fi
\providecommand\bibfield[2]{#2}
\providecommand\bibinfo[2]{#2}
\providecommand\natexlab[1]{#1}
\providecommand\showeprint[2][]{arXiv:#2}

\bibitem[Ajith et~al\mbox{.}(2024)]%
        {Ajith2024LitSearch}
\bibfield{author}{\bibinfo{person}{Anirudh Ajith}, \bibinfo{person}{Mengzhou Xia}, \bibinfo{person}{Alexis Chevalier}, \bibinfo{person}{Tanya Goyal}, \bibinfo{person}{Danqi Chen}, {and} \bibinfo{person}{Tianyu Gao}.} \bibinfo{year}{2024}\natexlab{}.
\newblock \showarticletitle{{L}it{S}earch: A Retrieval Benchmark for Scientific Literature Search}. In \bibinfo{booktitle}{\emph{Proceedings of the 2024 Conference on Empirical Methods in Natural Language Processing}}. \bibinfo{publisher}{Association for Computational Linguistics}, \bibinfo{address}{Miami, Florida, USA}, \bibinfo{pages}{15068--15083}.
\newblock
\href{https://doi.org/10.18653/v1/2024.emnlp-main.840}{doi:\nolinkurl{10.18653/v1/2024.emnlp-main.840}}


\bibitem[Ateia and Kruschwitz(2025)]%
        {Ateia2025BioRAGent}
\bibfield{author}{\bibinfo{person}{Samy Ateia} {and} \bibinfo{person}{Udo Kruschwitz}.} \bibinfo{year}{2025}\natexlab{}.
\newblock \showarticletitle{BioRAGent: A Retrieval-Augmented Generation System for Showcasing Generative Query Expansion and Domain-Specific Search for Scientific Q\&A}. In \bibinfo{booktitle}{\emph{Advances in Information Retrieval: 47th European Conference on Information Retrieval, ECIR 2025, Lucca, Italy, April 6–10, 2025, Proceedings, Part V}} (Lucca, Italy). \bibinfo{publisher}{Springer-Verlag}, \bibinfo{address}{Berlin, Heidelberg}, \bibinfo{pages}{1–5}.
\newblock
\showISBNx{978-3-031-88719-2}
\href{https://doi.org/10.1007/978-3-031-88720-8_1}{doi:\nolinkurl{10.1007/978-3-031-88720-8_1}}


\bibitem[Auer et~al\mbox{.}(2023)]%
        {Auer2023SciQA}
\bibfield{author}{\bibinfo{person}{S{\"o}ren Auer}, \bibinfo{person}{Dante A~C Barone}, \bibinfo{person}{Cassiano Bartz}, \bibinfo{person}{Eduardo~G Cortes}, \bibinfo{person}{Mohamad~Yaser Jaradeh}, \bibinfo{person}{Oliver Karras}, \bibinfo{person}{Manolis Koubarakis}, \bibinfo{person}{Dmitry Mouromtsev}, \bibinfo{person}{Dmitrii Pliukhin}, \bibinfo{person}{Daniil Radyush}, \bibinfo{person}{Ivan Shilin}, \bibinfo{person}{Markus Stocker}, {and} \bibinfo{person}{Eleni Tsalapati}.} \bibinfo{year}{2023}\natexlab{}.
\newblock \showarticletitle{The {SciQA} Scientific Question Answering Benchmark for Scholarly Knowledge}.
\newblock \bibinfo{journal}{\emph{Scientific Reports}} \bibinfo{volume}{13}, \bibinfo{number}{1} (\bibinfo{date}{May} \bibinfo{year}{2023}), \bibinfo{pages}{7240}.
\newblock
\href{https://doi.org/10.1038/s41598-023-33607-z}{doi:\nolinkurl{10.1038/s41598-023-33607-z}}


\bibitem[Berchansky et~al\mbox{.}(2024)]%
        {Berchansky2024CoTAR}
\bibfield{author}{\bibinfo{person}{Moshe Berchansky}, \bibinfo{person}{Daniel Fleischer}, \bibinfo{person}{Moshe Wasserblat}, {and} \bibinfo{person}{Peter Izsak}.} \bibinfo{year}{2024}\natexlab{}.
\newblock \showarticletitle{{C}o{TAR}: Chain-of-Thought Attribution Reasoning with Multi-level Granularity}. In \bibinfo{booktitle}{\emph{Findings of the Association for Computational Linguistics: EMNLP 2024}}. \bibinfo{publisher}{ACL}, \bibinfo{address}{Miami, Florida, USA}, \bibinfo{pages}{236--246}.
\newblock
\href{https://doi.org/10.18653/v1/2024.findings-emnlp.13}{doi:\nolinkurl{10.18653/v1/2024.findings-emnlp.13}}


\bibitem[Besrour and Färber(2025)]%
        {Besrour2025UnarXive}
\bibfield{author}{\bibinfo{person}{Ines Besrour} {and} \bibinfo{person}{Michael Färber}.} \bibinfo{year}{2025}\natexlab{}.
\newblock \bibinfo{title}{unarXive 2024}.
\newblock \bibinfo{howpublished}{\url{https://huggingface.co/datasets/ines-besrour/unarxive_2024}}.
\newblock
\newblock
\shownote{Accessed: 2025-06-18}.


\bibitem[Chang et~al\mbox{.}(2025)]%
        {Chang2024MAIN-RAG}
\bibfield{author}{\bibinfo{person}{Chia-Yuan Chang}, \bibinfo{person}{Zhimeng Jiang}, \bibinfo{person}{Vineeth Rakesh}, \bibinfo{person}{Menghai Pan}, \bibinfo{person}{Chin-Chia~Michael Yeh}, \bibinfo{person}{Guanchu Wang}, \bibinfo{person}{Mingzhi Hu}, \bibinfo{person}{Zhichao Xu}, \bibinfo{person}{Yan Zheng}, \bibinfo{person}{Mahashweta Das}, {and} \bibinfo{person}{Na Zou}.} \bibinfo{year}{2025}\natexlab{}.
\newblock \showarticletitle{{MAIN}-{RAG}: Multi-Agent Filtering Retrieval-Augmented Generation}. In \bibinfo{booktitle}{\emph{Proceedings of the 63rd Annual Meeting of the Association for Computational Linguistics (Volume 1: Long Papers)}}. \bibinfo{publisher}{Association for Computational Linguistics}, \bibinfo{address}{Vienna, Austria}, \bibinfo{pages}{2607--2622}.
\newblock
\href{https://doi.org/10.18653/v1/2025.acl-long.131}{doi:\nolinkurl{10.18653/v1/2025.acl-long.131}}


\bibitem[Dasigi et~al\mbox{.}(2021)]%
        {Dasigi2021Qasper}
\bibfield{author}{\bibinfo{person}{Pradeep Dasigi}, \bibinfo{person}{Kyle Lo}, \bibinfo{person}{Iz Beltagy}, \bibinfo{person}{Arman Cohan}, \bibinfo{person}{Noah~A. Smith}, {and} \bibinfo{person}{Matt Gardner}.} \bibinfo{year}{2021}\natexlab{}.
\newblock \showarticletitle{A Dataset of Information-Seeking Questions and Answers Anchored in Research Papers}. In \bibinfo{booktitle}{\emph{Proceedings of the 2021 Conference of the North American Chapter of the Association for Computational Linguistics: Human Language Technologies}}. \bibinfo{publisher}{ACL}, \bibinfo{address}{Online}, \bibinfo{pages}{4599--4610}.
\newblock
\href{https://doi.org/10.18653/v1/2021.naacl-main.365}{doi:\nolinkurl{10.18653/v1/2021.naacl-main.365}}


\bibitem[DeepEval(2025)]%
        {DeepEval2025}
\bibfield{author}{\bibinfo{person}{DeepEval}.} \bibinfo{year}{2025}\natexlab{}.
\newblock \bibinfo{title}{DeepEval: The Open-Source LLM Evaluation Framework}.
\newblock \bibinfo{howpublished}{\url{https://deepeval.com/}}.
\newblock
\newblock
\shownote{Accessed: 2025-06-18}.


\bibitem[Fierro et~al\mbox{.}(2024)]%
        {Fierro2024LearningToPlan}
\bibfield{author}{\bibinfo{person}{Constanza Fierro}, \bibinfo{person}{Reinald~Kim Amplayo}, \bibinfo{person}{Fantine Huot}, \bibinfo{person}{Nicola De~Cao}, \bibinfo{person}{Joshua Maynez}, \bibinfo{person}{Shashi Narayan}, {and} \bibinfo{person}{Mirella Lapata}.} \bibinfo{year}{2024}\natexlab{}.
\newblock \showarticletitle{Learning to Plan and Generate Text with Citations}. In \bibinfo{booktitle}{\emph{Proceedings of the 62nd Annual Meeting of the Association for Computational Linguistics (Volume 1: Long Papers)}}. \bibinfo{publisher}{ACL}, \bibinfo{address}{Bangkok, Thailand}, \bibinfo{pages}{11397--11417}.
\newblock
\href{https://doi.org/10.18653/v1/2024.acl-long.615}{doi:\nolinkurl{10.18653/v1/2024.acl-long.615}}


\bibitem[Gao et~al\mbox{.}(2023a)]%
        {Gao2023RARR}
\bibfield{author}{\bibinfo{person}{Luyu Gao}, \bibinfo{person}{Zhuyun Dai}, \bibinfo{person}{Panupong Pasupat}, \bibinfo{person}{Anthony Chen}, \bibinfo{person}{Arun~Tejasvi Chaganty}, \bibinfo{person}{Yicheng Fan}, \bibinfo{person}{Vincent Zhao}, \bibinfo{person}{Ni Lao}, \bibinfo{person}{Hongrae Lee}, \bibinfo{person}{Da-Cheng Juan}, {and} \bibinfo{person}{Kelvin Guu}.} \bibinfo{year}{2023}\natexlab{a}.
\newblock \showarticletitle{{RARR}: Researching and Revising What Language Models Say, Using Language Models}. In \bibinfo{booktitle}{\emph{Proceedings of the 61st Annual Meeting of the Association for Computational Linguistics (Volume 1: Long Papers)}}. \bibinfo{publisher}{ACL}, \bibinfo{address}{Toronto, Canada}, \bibinfo{pages}{16477--16508}.
\newblock
\href{https://doi.org/10.18653/v1/2023.acl-long.910}{doi:\nolinkurl{10.18653/v1/2023.acl-long.910}}


\bibitem[Gao et~al\mbox{.}(2023b)]%
        {Gao2023ALCE}
\bibfield{author}{\bibinfo{person}{Tianyu Gao}, \bibinfo{person}{Howard Yen}, \bibinfo{person}{Jiatong Yu}, {and} \bibinfo{person}{Danqi Chen}.} \bibinfo{year}{2023}\natexlab{b}.
\newblock \showarticletitle{Enabling Large Language Models to Generate Text with Citations}. In \bibinfo{booktitle}{\emph{Proceedings of the 2023 Conference on Empirical Methods in Natural Language Processing}}. \bibinfo{publisher}{ACL}, \bibinfo{address}{Singapore}, \bibinfo{pages}{6465--6488}.
\newblock
\href{https://doi.org/10.18653/v1/2023.emnlp-main.398}{doi:\nolinkurl{10.18653/v1/2023.emnlp-main.398}}


\bibitem[Huang and Chang(2024)]%
        {Huang2024Citation}
\bibfield{author}{\bibinfo{person}{Jie Huang} {and} \bibinfo{person}{Kevin Chang}.} \bibinfo{year}{2024}\natexlab{}.
\newblock \showarticletitle{Citation: A Key to Building Responsible and Accountable Large Language Models}. In \bibinfo{booktitle}{\emph{Findings of the Association for Computational Linguistics: NAACL 2024}}. \bibinfo{publisher}{ACL}, \bibinfo{address}{Mexico City, Mexico}, \bibinfo{pages}{464--473}.
\newblock
\href{https://doi.org/10.18653/v1/2024.findings-naacl.31}{doi:\nolinkurl{10.18653/v1/2024.findings-naacl.31}}


\bibitem[Huang et~al\mbox{.}(2025)]%
        {Huang2025Hallucination}
\bibfield{author}{\bibinfo{person}{Lei Huang}, \bibinfo{person}{Weijiang Yu}, \bibinfo{person}{Weitao Ma}, \bibinfo{person}{Weihong Zhong}, \bibinfo{person}{Zhangyin Feng}, \bibinfo{person}{Haotian Wang}, \bibinfo{person}{Qianglong Chen}, \bibinfo{person}{Weihua Peng}, \bibinfo{person}{Xiaocheng Feng}, \bibinfo{person}{Bing Qin}, {and} \bibinfo{person}{Ting Liu}.} \bibinfo{year}{2025}\natexlab{}.
\newblock \showarticletitle{A Survey on Hallucination in Large Language Models: Principles, Taxonomy, Challenges, and Open Questions}.
\newblock \bibinfo{journal}{\emph{ACM Trans. Inf. Syst.}} \bibinfo{volume}{43}, \bibinfo{number}{2}, Article \bibinfo{articleno}{42} (\bibinfo{date}{Jan.} \bibinfo{year}{2025}), \bibinfo{numpages}{55}~pages.
\newblock
\href{https://doi.org/10.1145/3703155}{doi:\nolinkurl{10.1145/3703155}}


\bibitem[Jang and Li(2024)]%
        {Jang2024AU-RAG}
\bibfield{author}{\bibinfo{person}{Jisoo Jang} {and} \bibinfo{person}{Wen-Syan Li}.} \bibinfo{year}{2024}\natexlab{}.
\newblock \showarticletitle{AU-RAG: Agent-based Universal Retrieval Augmented Generation}. In \bibinfo{booktitle}{\emph{Proceedings of the 2024 Annual International ACM SIGIR Conference on Research and Development in Information Retrieval in the Asia Pacific Region}} (Tokyo, Japan) \emph{(\bibinfo{series}{SIGIR-AP 2024})}. \bibinfo{publisher}{ACM}, \bibinfo{address}{New York, NY, USA}, \bibinfo{pages}{2–11}.
\newblock
\href{https://doi.org/10.1145/3673791.3698416}{doi:\nolinkurl{10.1145/3673791.3698416}}


\bibitem[Jin et~al\mbox{.}(2019)]%
        {Jin2019PubMedQA}
\bibfield{author}{\bibinfo{person}{Qiao Jin}, \bibinfo{person}{Bhuwan Dhingra}, \bibinfo{person}{Zhengping Liu}, \bibinfo{person}{William Cohen}, {and} \bibinfo{person}{Xinghua Lu}.} \bibinfo{year}{2019}\natexlab{}.
\newblock \showarticletitle{{P}ub{M}ed{QA}: A Dataset for Biomedical Research Question Answering}. In \bibinfo{booktitle}{\emph{Proceedings of the 2019 Conference on Empirical Methods in Natural Language Processing and the 9th International Joint Conference on Natural Language Processing (EMNLP-IJCNLP)}}. \bibinfo{publisher}{ACL}, \bibinfo{address}{Hong Kong, China}, \bibinfo{pages}{2567--2577}.
\newblock
\href{https://doi.org/10.18653/v1/D19-1259}{doi:\nolinkurl{10.18653/v1/D19-1259}}


\bibitem[Khattab et~al\mbox{.}(2021)]%
        {khattab2021relevance}
\bibfield{author}{\bibinfo{person}{Omar Khattab}, \bibinfo{person}{Christopher Potts}, {and} \bibinfo{person}{Matei Zaharia}.} \bibinfo{year}{2021}\natexlab{}.
\newblock \showarticletitle{Relevance-guided Supervision for {O}pen{QA} with {C}ol{BERT}}.
\newblock \bibinfo{journal}{\emph{Transactions of the Association for Computational Linguistics}}  \bibinfo{volume}{9} (\bibinfo{year}{2021}), \bibinfo{pages}{929--944}.
\newblock
\href{https://doi.org/10.1162/tacl_a_00405}{doi:\nolinkurl{10.1162/tacl_a_00405}}


\bibitem[Krithara et~al\mbox{.}(2023)]%
        {Krithara2023BioASQ-QA}
\bibfield{author}{\bibinfo{person}{Anastasia Krithara}, \bibinfo{person}{Anastasios Nentidis}, \bibinfo{person}{Konstantinos Bougiatiotis}, {and} \bibinfo{person}{Georgios Paliouras}.} \bibinfo{year}{2023}\natexlab{}.
\newblock \showarticletitle{BioASQ-QA: A manually curated corpus for Biomedical Question Answering}.
\newblock \bibinfo{journal}{\emph{Scientific Data}} \bibinfo{volume}{10}, \bibinfo{number}{1} (\bibinfo{date}{27 Mar} \bibinfo{year}{2023}), \bibinfo{pages}{170}.
\newblock
\href{https://doi.org/10.1038/s41597-023-02068-4}{doi:\nolinkurl{10.1038/s41597-023-02068-4}}


\bibitem[Lewis et~al\mbox{.}(2020)]%
        {Lewis2020RAG}
\bibfield{author}{\bibinfo{person}{Patrick Lewis}, \bibinfo{person}{Ethan Perez}, \bibinfo{person}{Aleksandra Piktus}, \bibinfo{person}{Fabio Petroni}, \bibinfo{person}{Vladimir Karpukhin}, \bibinfo{person}{Naman Goyal}, \bibinfo{person}{Heinrich K\"{u}ttler}, \bibinfo{person}{Mike Lewis}, \bibinfo{person}{Wen-tau Yih}, \bibinfo{person}{Tim Rockt\"{a}schel}, \bibinfo{person}{Sebastian Riedel}, {and} \bibinfo{person}{Douwe Kiela}.} \bibinfo{year}{2020}\natexlab{}.
\newblock \showarticletitle{Retrieval-Augmented Generation for Knowledge-Intensive NLP Tasks}. In \bibinfo{booktitle}{\emph{Advances in Neural Information Processing Systems}}, Vol.~\bibinfo{volume}{33}. \bibinfo{publisher}{Curran Associates, Inc.}, \bibinfo{address}{Red Hook, NY, USA}, \bibinfo{pages}{9459--9474}.
\newblock
\urldef\tempurl%
\url{https://dl.acm.org/doi/abs/10.5555/3495724.3496517}
\showURL{%
\tempurl}


\bibitem[Liu et~al\mbox{.}(2023)]%
        {Liu2023G-Eval}
\bibfield{author}{\bibinfo{person}{Yang Liu}, \bibinfo{person}{Dan Iter}, \bibinfo{person}{Yichong Xu}, \bibinfo{person}{Shuohang Wang}, \bibinfo{person}{Ruochen Xu}, {and} \bibinfo{person}{Chenguang Zhu}.} \bibinfo{year}{2023}\natexlab{}.
\newblock \showarticletitle{{G}-Eval: {NLG} Evaluation using Gpt-4 with Better Human Alignment}. In \bibinfo{booktitle}{\emph{Proceedings of the 2023 Conference on Empirical Methods in Natural Language Processing}}. \bibinfo{publisher}{ACL}, \bibinfo{address}{Singapore}, \bibinfo{pages}{2511--2522}.
\newblock
\href{https://doi.org/10.18653/v1/2023.emnlp-main.153}{doi:\nolinkurl{10.18653/v1/2023.emnlp-main.153}}


\bibitem[Lu et~al\mbox{.}(2022)]%
        {Pan2022ScienceQA}
\bibfield{author}{\bibinfo{person}{Pan Lu}, \bibinfo{person}{Swaroop Mishra}, \bibinfo{person}{Tony Xia}, \bibinfo{person}{Liang Qiu}, \bibinfo{person}{Kai-Wei Chang}, \bibinfo{person}{Song-Chun Zhu}, \bibinfo{person}{Oyvind Tafjord}, \bibinfo{person}{Peter Clark}, {and} \bibinfo{person}{Ashwin Kalyan}.} \bibinfo{year}{2022}\natexlab{}.
\newblock \showarticletitle{Learn to explain: multimodal reasoning via thought chains for science question answering}. In \bibinfo{booktitle}{\emph{Proceedings of the 36th International Conference on Neural Information Processing Systems}} (New Orleans, LA, USA) \emph{(\bibinfo{series}{NIPS '22})}. \bibinfo{publisher}{Curran Associates Inc.}, \bibinfo{address}{Red Hook, NY, USA}, Article \bibinfo{articleno}{182}, \bibinfo{numpages}{15}~pages.
\newblock
\urldef\tempurl%
\url{https://dl.acm.org/doi/10.5555/3600270.3600452}
\showURL{%
\tempurl}


\bibitem[Lála et~al\mbox{.}(2023)]%
        {Lála2023PaperQA}
\bibfield{author}{\bibinfo{person}{Jakub Lála}, \bibinfo{person}{Odhran O'Donoghue}, \bibinfo{person}{Aleksandar Shtedritski}, \bibinfo{person}{Sam Cox}, \bibinfo{person}{Samuel~G. Rodriques}, {and} \bibinfo{person}{Andrew~D. White}.} \bibinfo{year}{2023}\natexlab{}.
\newblock \bibinfo{title}{PaperQA: Retrieval-Augmented Generative Agent for Scientific Research}.
\newblock
\showeprint[arxiv]{2312.07559}~[cs.CL]


\bibitem[Mosquera et~al\mbox{.}(2025)]%
        {rayo2025hybrid}
\bibfield{author}{\bibinfo{person}{Jhon Stewar~Rayo Mosquera}, \bibinfo{person}{Carlos Raúl De La~Rosa Peredo}, {and} \bibinfo{person}{Mario~Garrido Córdoba}.} \bibinfo{year}{2025}\natexlab{}.
\newblock \showarticletitle{A Hybrid Approach to Information Retrieval and Answer Generation for Regulatory Texts}. In \bibinfo{booktitle}{\emph{Proceedings of the 1st Regulatory NLP Workshop (RegNLP 2025)}}. \bibinfo{publisher}{ACL}, \bibinfo{address}{Abu Dhabi, UAE}, \bibinfo{pages}{31--35}.
\newblock
\urldef\tempurl%
\url{https://aclanthology.org/2025.regnlp-1.5/}
\showURL{%
\tempurl}


\bibitem[Nogueira et~al\mbox{.}(2019)]%
        {nogueira2020document}
\bibfield{author}{\bibinfo{person}{Rodrigo Nogueira}, \bibinfo{person}{Wei Yang}, \bibinfo{person}{Jimmy Lin}, {and} \bibinfo{person}{Kyunghyun Cho}.} \bibinfo{year}{2019}\natexlab{}.
\newblock \bibinfo{title}{Document Expansion by Query Prediction}.
\newblock
\showeprint[arxiv]{1904.08375}~[cs.IR]


\bibitem[Ramu et~al\mbox{.}(2024)]%
        {Ramu2024PostHoc}
\bibfield{author}{\bibinfo{person}{Pritika Ramu}, \bibinfo{person}{Koustava Goswami}, \bibinfo{person}{Apoorv Saxena}, {and} \bibinfo{person}{Balaji~Vasan Srinivasan}.} \bibinfo{year}{2024}\natexlab{}.
\newblock \showarticletitle{Enhancing Post-Hoc Attributions in Long Document Comprehension via Coarse Grained Answer Decomposition}. In \bibinfo{booktitle}{\emph{Proceedings of the 2024 Conference on Empirical Methods in Natural Language Processing}}. \bibinfo{publisher}{ACL}, \bibinfo{address}{Miami, Florida, USA}, \bibinfo{pages}{17790--17806}.
\newblock
\href{https://doi.org/10.18653/v1/2024.emnlp-main.985}{doi:\nolinkurl{10.18653/v1/2024.emnlp-main.985}}


\bibitem[Robertson and Walker(1994)]%
        {Robertson1994BM25}
\bibfield{author}{\bibinfo{person}{S.~E. Robertson} {and} \bibinfo{person}{S. Walker}.} \bibinfo{year}{1994}\natexlab{}.
\newblock \showarticletitle{Some simple effective approximations to the 2-Poisson model for probabilistic weighted retrieval}. In \bibinfo{booktitle}{\emph{ACM SIGIR'94}} (Dublin, Ireland). \bibinfo{publisher}{Springer-Verlag}, \bibinfo{address}{Berlin, Heidelberg}, \bibinfo{pages}{232–241}.
\newblock
\showISBNx{038719889X}
\href{https://doi.org/10.1007/978-1-4471-2099-5_24}{doi:\nolinkurl{10.1007/978-1-4471-2099-5_24}}


\bibitem[Robertson et~al\mbox{.}(1994)]%
        {Robertson1994Okapi}
\bibfield{author}{\bibinfo{person}{Stephen~E. Robertson}, \bibinfo{person}{Steve Walker}, \bibinfo{person}{Susan Jones}, \bibinfo{person}{Micheline Hancock{-}Beaulieu}, {and} \bibinfo{person}{Mike Gatford}.} \bibinfo{year}{1994}\natexlab{}.
\newblock \showarticletitle{Okapi at {TREC-3}}. In \bibinfo{booktitle}{\emph{TREC'94}}, Vol.~\bibinfo{volume}{500-225}. \bibinfo{publisher}{NIST}, \bibinfo{address}{Gaithersburg, USA}, \bibinfo{pages}{109--126}.
\newblock
\urldef\tempurl%
\url{http://trec.nist.gov/pubs/trec3/papers/city.ps.gz}
\showURL{%
\tempurl}


\bibitem[Saeid and Kopinski(2024)]%
        {Saeid2024AgentFusion}
\bibfield{author}{\bibinfo{person}{Yasser Saeid} {and} \bibinfo{person}{Thomas Kopinski}.} \bibinfo{year}{2024}\natexlab{}.
\newblock \showarticletitle{AgentFusion: A Multi-Agent Approach to Accurate Text Generation}. In \bibinfo{booktitle}{\emph{2024 International Conference on Electrical and Computer Engineering Researches (ICECER)}}. \bibinfo{publisher}{IEEE}, \bibinfo{address}{Gaborone, Botswana}, \bibinfo{pages}{1--8}.
\newblock
\href{https://doi.org/10.1109/ICECER62944.2024.10920460}{doi:\nolinkurl{10.1109/ICECER62944.2024.10920460}}


\bibitem[Saier and F\"{a}rber(2020)]%
        {Saier2020unarXive}
\bibfield{author}{\bibinfo{person}{Tarek Saier} {and} \bibinfo{person}{Michael F\"{a}rber}.} \bibinfo{year}{2020}\natexlab{}.
\newblock \showarticletitle{unarXive: a large scholarly data set with publications’ full-text, annotated in-text citations, and links to metadata}.
\newblock \bibinfo{journal}{\emph{Scientometrics}} \bibinfo{volume}{125}, \bibinfo{number}{3} (\bibinfo{date}{Dec.} \bibinfo{year}{2020}), \bibinfo{pages}{3085–3108}.
\newblock
\showISSN{0138-9130}
\href{https://doi.org/10.1007/s11192-020-03382-z}{doi:\nolinkurl{10.1007/s11192-020-03382-z}}


\bibitem[Saier et~al\mbox{.}(2024)]%
        {Saier2024unarxive}
\bibfield{author}{\bibinfo{person}{Tarek Saier}, \bibinfo{person}{Johan Krause}, {and} \bibinfo{person}{Michael F\"{a}rber}.} \bibinfo{year}{2024}\natexlab{}.
\newblock \showarticletitle{unarXive 2022: All arXiv Publications Pre-Processed for NLP, Including Structured Full-Text and Citation Network}. In \bibinfo{booktitle}{\emph{Proceedings of the 2023 ACM/IEEE Joint Conference on Digital Libraries}} \emph{(\bibinfo{series}{JCDL '23})}. \bibinfo{publisher}{IEEE Press}, \bibinfo{address}{Santa Fe, USA}, \bibinfo{pages}{66–70}.
\newblock
\showISBNx{9798350399318}
\href{https://doi.org/10.1109/JCDL57899.2023.00020}{doi:\nolinkurl{10.1109/JCDL57899.2023.00020}}


\bibitem[Schreieder et~al\mbox{.}(2025)]%
        {Schreieder2025AttributionCitationQuotation}
\bibfield{author}{\bibinfo{person}{Tobias Schreieder}, \bibinfo{person}{Tim Schopf}, {and} \bibinfo{person}{Michael Färber}.} \bibinfo{year}{2025}\natexlab{}.
\newblock \bibinfo{title}{Attribution, Citation, and Quotation: A Survey of Evidence-based Text Generation with Large Language Models}.
\newblock
\showeprint[arxiv]{2508.15396}~[cs.CL]


\bibitem[Singhal et~al\mbox{.}(2025)]%
        {Singhal2025QA}
\bibfield{author}{\bibinfo{person}{Karan Singhal}, \bibinfo{person}{Tao Tu}, \bibinfo{person}{Juraj Gottweis}, \bibinfo{person}{Rory Sayres}, \bibinfo{person}{Ellery Wulczyn}, \bibinfo{person}{Mohamed Amin}, \bibinfo{person}{Le Hou}, \bibinfo{person}{Kevin Clark}, \bibinfo{person}{Stephen~R. Pfohl}, \bibinfo{person}{Heather Cole-Lewis}, \bibinfo{person}{Darlene Neal}, \bibinfo{person}{Qazi~Mamunur Rashid}, \bibinfo{person}{Mike Schaekermann}, \bibinfo{person}{Amy Wang}, \bibinfo{person}{Dev Dash}, \bibinfo{person}{Jonathan~H. Chen}, \bibinfo{person}{Nigam~H. Shah}, \bibinfo{person}{Sami Lachgar}, \bibinfo{person}{Philip~Andrew Mansfield}, \bibinfo{person}{Sushant Prakash}, \bibinfo{person}{Bradley Green}, \bibinfo{person}{Ewa Dominowska}, \bibinfo{person}{Blaise Ag{\"u}era~y Arcas}, \bibinfo{person}{Nenad Toma{\v{s}}ev}, \bibinfo{person}{Yun Liu}, \bibinfo{person}{Renee Wong}, \bibinfo{person}{Christopher Semturs}, \bibinfo{person}{S.~Sara Mahdavi}, \bibinfo{person}{Joelle~K. Barral}, \bibinfo{person}{Dale~R.
  Webster}, \bibinfo{person}{Greg~S. Corrado}, \bibinfo{person}{Yossi Matias}, \bibinfo{person}{Shekoofeh Azizi}, \bibinfo{person}{Alan Karthikesalingam}, {and} \bibinfo{person}{Vivek Natarajan}.} \bibinfo{year}{2025}\natexlab{}.
\newblock \showarticletitle{Toward expert-level medical question answering with large language models}.
\newblock \bibinfo{journal}{\emph{Nature Medicine}} \bibinfo{volume}{31}, \bibinfo{number}{3} (\bibinfo{date}{01 Mar} \bibinfo{year}{2025}), \bibinfo{pages}{943--950}.
\newblock
\showISSN{1546-170X}
\href{https://doi.org/10.1038/s41591-024-03423-7}{doi:\nolinkurl{10.1038/s41591-024-03423-7}}


\bibitem[Taylor et~al\mbox{.}(2022)]%
        {Taylor2022Galactica}
\bibfield{author}{\bibinfo{person}{Ross Taylor}, \bibinfo{person}{Marcin Kardas}, \bibinfo{person}{Guillem Cucurull}, \bibinfo{person}{Thomas Scialom}, \bibinfo{person}{Anthony Hartshorn}, \bibinfo{person}{Elvis Saravia}, \bibinfo{person}{Andrew Poulton}, \bibinfo{person}{Viktor Kerkez}, {and} \bibinfo{person}{Robert Stojnic}.} \bibinfo{year}{2022}\natexlab{}.
\newblock \bibinfo{title}{Galactica: A Large Language Model for Science}.
\newblock
\showeprint[arxiv]{2211.09085}~[cs.CL]


\bibitem[Wang et~al\mbox{.}(2024)]%
        {Wang2024E5}
\bibfield{author}{\bibinfo{person}{Liang Wang}, \bibinfo{person}{Nan Yang}, \bibinfo{person}{Xiaolong Huang}, \bibinfo{person}{Binxing Jiao}, \bibinfo{person}{Linjun Yang}, \bibinfo{person}{Daxin Jiang}, \bibinfo{person}{Rangan Majumder}, {and} \bibinfo{person}{Furu Wei}.} \bibinfo{year}{2024}\natexlab{}.
\newblock \bibinfo{title}{Text Embeddings by Weakly-Supervised Contrastive Pre-training}.
\newblock
\showeprint[arxiv]{2212.03533}~[cs.CL]


\bibitem[Wei et~al\mbox{.}(2022)]%
        {Wei2022CoT}
\bibfield{author}{\bibinfo{person}{Jason Wei}, \bibinfo{person}{Xuezhi Wang}, \bibinfo{person}{Dale Schuurmans}, \bibinfo{person}{Maarten Bosma}, \bibinfo{person}{Brian Ichter}, \bibinfo{person}{Fei Xia}, \bibinfo{person}{Ed~H. Chi}, \bibinfo{person}{Quoc~V. Le}, {and} \bibinfo{person}{Denny Zhou}.} \bibinfo{year}{2022}\natexlab{}.
\newblock \showarticletitle{Chain-of-thought prompting elicits reasoning in large language models}. In \bibinfo{booktitle}{\emph{Proceedings of the 36th International Conference on Neural Information Processing Systems}} (New Orleans, LA, USA) \emph{(\bibinfo{series}{NIPS '22})}. \bibinfo{publisher}{Curran Associates Inc.}, \bibinfo{address}{Red Hook, NY, USA}, Article \bibinfo{articleno}{1800}, \bibinfo{numpages}{14}~pages.
\newblock
\showISBNx{9781713871088}
\urldef\tempurl%
\url{https://dl.acm.org/doi/10.5555/3600270.3602070}
\showURL{%
\tempurl}


\bibitem[Wolfson et~al\mbox{.}(2020)]%
        {wolfson2020break}
\bibfield{author}{\bibinfo{person}{Tomer Wolfson}, \bibinfo{person}{Mor Geva}, \bibinfo{person}{Ankit Gupta}, \bibinfo{person}{Matt Gardner}, \bibinfo{person}{Yoav Goldberg}, \bibinfo{person}{Daniel Deutch}, {and} \bibinfo{person}{Jonathan Berant}.} \bibinfo{year}{2020}\natexlab{}.
\newblock \showarticletitle{Break It Down: A Question Understanding Benchmark}.
\newblock \bibinfo{journal}{\emph{Transactions of the Association for Computational Linguistics}}  \bibinfo{volume}{8} (\bibinfo{year}{2020}), \bibinfo{pages}{183--198}.
\newblock
\href{https://doi.org/10.1162/tacl_a_00309}{doi:\nolinkurl{10.1162/tacl_a_00309}}


\bibitem[Yang et~al\mbox{.}(2024)]%
        {Yang2024IM-RAG}
\bibfield{author}{\bibinfo{person}{Diji Yang}, \bibinfo{person}{Jinmeng Rao}, \bibinfo{person}{Kezhen Chen}, \bibinfo{person}{Xiaoyuan Guo}, \bibinfo{person}{Yawen Zhang}, \bibinfo{person}{Jie Yang}, {and} \bibinfo{person}{Yi Zhang}.} \bibinfo{year}{2024}\natexlab{}.
\newblock \showarticletitle{IM-RAG: Multi-Round Retrieval-Augmented Generation Through Learning Inner Monologues}. In \bibinfo{booktitle}{\emph{Proceedings of the 47th International ACM SIGIR Conference on Research and Development in Information Retrieval}} (Washington DC, USA) \emph{(\bibinfo{series}{SIGIR '24})}. \bibinfo{publisher}{ACM}, \bibinfo{address}{New York, NY, USA}, \bibinfo{pages}{730–740}.
\newblock
\showISBNx{9798400704314}
\href{https://doi.org/10.1145/3626772.3657760}{doi:\nolinkurl{10.1145/3626772.3657760}}


\bibitem[Zhang et~al\mbox{.}(2024)]%
        {Zhang2024Summarization}
\bibfield{author}{\bibinfo{person}{Tianyi Zhang}, \bibinfo{person}{Faisal Ladhak}, \bibinfo{person}{Esin Durmus}, \bibinfo{person}{Percy Liang}, \bibinfo{person}{Kathleen McKeown}, {and} \bibinfo{person}{Tatsunori~B. Hashimoto}.} \bibinfo{year}{2024}\natexlab{}.
\newblock \showarticletitle{Benchmarking Large Language Models for News Summarization}.
\newblock \bibinfo{journal}{\emph{Transactions of the Association for Computational Linguistics}}  \bibinfo{volume}{12} (\bibinfo{year}{2024}), \bibinfo{pages}{39--57}.
\newblock
\href{https://doi.org/10.1162/tacl_a_00632}{doi:\nolinkurl{10.1162/tacl_a_00632}}


\bibitem[Zhao et~al\mbox{.}(2024)]%
        {Zhao2024Longagent}
\bibfield{author}{\bibinfo{person}{Jun Zhao}, \bibinfo{person}{Can Zu}, \bibinfo{person}{Xu Hao}, \bibinfo{person}{Yi Lu}, \bibinfo{person}{Wei He}, \bibinfo{person}{Yiwen Ding}, \bibinfo{person}{Tao Gui}, \bibinfo{person}{Qi Zhang}, {and} \bibinfo{person}{Xuanjing Huang}.} \bibinfo{year}{2024}\natexlab{}.
\newblock \showarticletitle{{LONGAGENT}: Achieving Question Answering for 128k-Token-Long Documents through Multi-Agent Collaboration}. In \bibinfo{booktitle}{\emph{Proceedings of the 2024 Conference on Empirical Methods in Natural Language Processing}}. \bibinfo{publisher}{ACL}, \bibinfo{address}{Miami, Florida, USA}, \bibinfo{pages}{16310--16324}.
\newblock
\href{https://doi.org/10.18653/v1/2024.emnlp-main.912}{doi:\nolinkurl{10.18653/v1/2024.emnlp-main.912}}


\end{thebibliography}
